\documentstyle[prd,aps]{revtex}

\newcommand{\bea}{\begin{eqnarray}}
\newcommand{\eea}{\end{eqnarray}}

\begin{document}
%%%%%%%%%%%%%%%%%%%%%%%%%%%%%%%%%%%%%%%%%%%%%%%%%%%%%%%%%%%%%%%
\draft
%  For 2 column format.
\twocolumn[\hsize\textwidth\columnwidth\hsize\csname
@twocolumnfalse\endcsname

%%%%%%%%%%%%%%%%%%%%%%%%%%%%%%%%%%%%%%%%%%%%%%%%%%%%%%%%%%%%%%%
\title{Inflationary spectra in generalized gravity: Unified forms}
\author{ Hyerim Noh${}^{(a,b)}$ and Jai-chan Hwang${}^{(c,b)}$ \\
        ${}^{(a)}$ Korea Astronomy Observatory, Daejon, Korea \\
        ${}^{(b)}$ Institute of Astronomy, Madingley Road, Cambridge, UK \\
        ${}^{(c)}$ Department of Astronomy and Atmospheric Sciences,
                   Kyungpook National University, Taegu, Korea }
\date{\today}
\maketitle

%%%%%%%%%%%%%%%%%%%%%%%%%%%%%%%%%%%%%%%%%%%%%%%%%%%%%%%%%%%%%%%
\begin{abstract}

The classical evolution and the quantum generation processes 
of the scalar- and tensor-type cosmological perturbations in the 
context of a broad class of generalized gravity theories are 
presented in unified forms.
The exact forms of final spectra of the two types of structures 
generated during a generalized slow-roll inflation are derived.
Results in generalized gravity are characterized by two additional
parameters which are the coupling between gravity and field 
$f(\phi,R)$, and the nonminimal coupling in the kinetic part of 
the field $\omega(\phi)$.
Our general results include widely studied gravity theories and
inflation models as special cases, and show how the well known
consistency relation and spectra in ordinary Einstein gravity inflation
models are affected by the generalized nature of the gravity theories.

\end{abstract}

%%%%%%%%%%%%%%%%%%%%%%%%%%%%%%%%%%%%%%%%%%%%%%%%%%%%%%%%%%%%%%%
%  For 2 column format.
\vskip2pc]

%%%%%%%%%%%%%%%%%%%%%%%%%%%%%%%%%%%%%%%%%%%%%%%%%%%%%%%%%%%%%%%
\section{Introduction}
                                       \label{sec:Introduction}

Lifshitz instability theory \cite{Lifshitz-1946}, the relativistic 
linear perturbation theory of an expanding Friedmann world model, 
first presented in 1946 has been studied in the literature over 
more than a half century 
\cite{perturbation,Ford-Parker-1977,Lukash-1980,Mukhanov-1988}.
The observed cosmological structures in the large-scale and
in the early universe are generally believed to behave as small
deviations from the homogeneous and isotropic background world model.
Under such a situation the relativisitic cosmological perturbation 
analysis becomes manageable due to the assumed linearity of the structures.
Recent observational advances of the CMBR anisotropies conform/reinforce
the validity of the two basic assumptions used in most
of the cosmological structure formation theories: the homogeous and
isotropic Friedmann world model, and the linearity of the
imposed structures.

However, the observational evidences do not necessarily
constrain the underlying gravity theory, especially during the
seed generating stage in the very early univese, to be Einstein one.
Generalized forms of gravity appear ubiquitously in any reasonable
attempts to understand the quantum aspects of the gravity theory, and
also naturally appear in the low energy limits of diverse attempts to
unify gravity with other fundamental forces, like the Kaluza-Klein, 
the supergravity, the string/M-theory programs.
Modifying terms appear naturally in the quantization processes of 
the gravity theory in a way toward the quantum gravity.
Thus, there arises a growing chance that the early stages of the universe
were governed by the gravity more general than Einstein one.

Reflecting such possibilities, there have been many studies of
the world models as well as the perturbations based on variety of 
generalized gravity theories \cite{Nariai-1973,SBB-1989,MFB-1992};
for our study see \cite{H-GGT,HN-GGT-1996,Hwang-CT-1997,Hwang-GW-1998}.
In this paper we will present
the classical evolution and quantum generation processes, and the
consequent inflationary spectra in unified forms which include
(1) the scalar- and the tensor-type structures, and
(2) the fluid and the field in Einstein gravity, and the field
in a class of generalized gravity theories.

We set $c \equiv 1$.

%%%%%%%%%%%%%%%%%%%%%%%%%%%%%%%%%%%%%%%%%%%%%%%%%%%%%%%%%%%%%%
\section{Gravity and world model}
                                       \label{sec:GGT}

We consider gravity theories with the following action
\bea
   & & S = \int d^4 x \sqrt{-g} \Big[ {1 \over 2} f (\phi, R)
       - {1\over 2} \omega (\phi) \phi^{;a} \phi_{,a} - V(\phi)
   \nonumber \\
   & & \qquad \qquad \qquad \qquad
       + L_m \Big],
   \label{action}
\eea
where $f(\phi, R)$ is a general algebraic function of the scalar field $\phi$
and the scalar curvature $R$; $\omega (\phi)$ and $V (\phi)$ are
general algebraic functions of $\phi$.
$L_m$ is the matter Lagrangian with the hydrodynamic 
energy-momentum tensor $T_{ab}$ defined as $\delta ( \sqrt{-g} L_m )
\equiv {1 \over 2} \sqrt{-g} T^{ab} \delta g_{ab}$.
Our generalized gravity includes as subset \cite{HN-GGT-1996}:
$f(R)$ gravity which includes $R^2$ gravity,
the scalar-tensor theory which includes the Jordan-Brans-Dicke theory
\cite{JBD}, the non-minimally coupled scalar field,
the induced gravity \cite{induced},
the low-energy effective action of string theory \cite{string}, etc.
It does not, however, include higher-derivative theories with
terms like $R^{ab} R_{ab}$, see \cite{NH-Rab}. 

We consider a spatially homogeneous and isotropic Friedmann world model with 
the most general spacetime dependent perturbations
\bea
   & & d s^2 = - \left( 1 + 2 \alpha \right) d t^2
       - 2 a \left( \beta_{,\alpha} +B_\alpha \right) d t d x^\alpha
   \nonumber \\
   & & \quad
       + a^2 \Big[ g_{\alpha\beta}^{(3)} \left( 1 + 2 \varphi \right)
       + 2 \gamma_{,\alpha|\beta} 
       + 2 C_{(\alpha|\beta)} + 2 C_{\alpha\beta}
       \Big] d x^\alpha d x^\beta.
   \nonumber \\
\eea
$\alpha$, $\beta$, $\gamma$, and $\varphi$ indicate the scalar-type structure;
the transverse $B_\alpha$ and $C_\alpha$ indicate the vector-type structure;
the transverse-tracefree $C_{\alpha\beta}$ indicates the tensor-type structure.
The three types of structures are related to the density condensation,
the rotation, and the gravitational wave, respectively.
Since, to the linear order in the Friedmann background, these three types 
of structures evolve independently, we can handle them separately.
Indices of $B_\alpha$, $C_\alpha$ and $C_{\alpha\beta}$ are based on 
$g_{\alpha\beta}^{(3)}$; a vertical bar indicates a covariant derivative
based on $g_{\alpha\beta}^{(3)}$.

We also consider the general perturbations in the 
hydrodynamic energy-momentum tensor and the scalar field
$T_{ab} ({\bf x}, t) = \bar T_{ab} (t) + \delta T_{ab} ({\bf x}, t)$ and
$\phi ({\bf x}, t) = \bar \phi (t) + \delta \phi ({\bf x}, t)$.
The perturbed order energy-momentum tensor in terms of the hydrodynamic fluid
quantities is
\bea
   & & T^0_0 = - \left( \bar \mu + \delta \mu \right), \quad
       T^0_\alpha = \left( \mu + p \right) v_\alpha,
   \nonumber \\
   & & T^\alpha_\beta = \left( \bar p + \delta p \right)
       \delta^\alpha_\beta + \pi^\alpha_\beta,
   \label{Tab}
\eea
where $v_\alpha$ and $\pi^\alpha_\beta$ are based on $g_{\alpha\beta}^{(3)}$.

We introduce the following gauge-invariant combinations
\bea
   & & \varphi_v \equiv \varphi - {aH \over k} v, \quad
       \varphi_{\delta \phi} \equiv \varphi - {H \over \dot \phi} \delta \phi
       \equiv - {H \over \dot \phi} \delta \phi_\varphi,
   \nonumber \\
   & & \varphi_\chi \equiv \varphi - H \chi,
   \label{varphi-delta-phi}
\eea
where $k$ is a comoving wavenumber and $H \equiv \dot a/a$;
an overdot and a prime indicate time derivatives based on $t$ and
the conformal time $\eta$, respectively, with $dt \equiv a d \eta$. 
$\varphi_\chi$ is the same as $\varphi$ in the zero-shear gauge 
which sets $\chi \equiv a (\beta + a \dot \gamma)$ equals to zero 
as the gauge condition;
$\varphi_v$ is the same as $\varphi$ in the comoving gauge condition
which takes $v/k \equiv 0$ as the gauge condition; $v$ introduced as 
$v_\alpha \equiv - v_{,\alpha}/k$ is a velocity related 
scalar-type perturbation variable.
{}For the scalar field the velocity related effective fluid quantity
becomes $a (\mu + p) v / k = \dot \phi \delta \phi$, 
thus the uniform-field gauge with $\delta \phi \equiv 0$ coincides with
the comoving gauge condition \cite{BST-1983}.
The gauge-invariant combination $\varphi_v$ was first introduced by 
Lukash in 1980 \cite{Lukash-1980}; in the following we will notice the 
profound importance of $\varphi_v$, the Lukash variable, in handling
the scalar-type cosmological perturbations.

The equations for background are:
\bea
   & & H^2 = {1 \over 3F} \left[ \mu
       + {1 \over 2} \left( \omega \dot \phi^2
       - f + RF + 2V \right) - 3 H \dot F \right]
       - {K \over a^2},
   \nonumber \\
   \label{BG1} \\
   & & \dot H = - {1\over 2 F} \left( \mu + p + \omega \dot \phi^2
       + \ddot F - H \dot F \right) + {K \over a^2},
   \label{BG2} \\
   & & R = 6 \Big( 2 H^2 + \dot H + {K \over a^2} \Big),
   \label{BG3} \\
   & & \ddot \phi + 3 H \dot \phi
       + {1 \over 2 \omega} \left( \omega_{,\phi} \dot \phi^2
       - f_{,\phi} + 2 V_{,\phi} \right) = 0,
   \label{BG4} \\
   & & \dot \mu + 3 H \left( \mu + p \right) = 0,
   \label{BG5}
\eea
where $F \equiv \partial f / (\partial R)$.
Equation (\ref{BG2}) follows from the rest of the equations.
$K$ is the sign of the background spatial curvature. 
In Einstein gravity limit we have $F = 1/(8 \pi G)$.
Our gravity theory includes the cosmological constant, 
$\Lambda$\footnote{
         It can be simulated using either the scalar field or the fluid.
         Using the scalar field we let $V \rightarrow V + \Lambda/(8 \pi G)$.
         Using the fluid we let $\mu \rightarrow \mu + \Lambda/(8 \pi G)$ and
         $p \rightarrow p - \Lambda/(8 \pi G)$.
         This causes a change only in eq. (\ref{BG1}).
         }.

%%%%%%%%%%%%%%%%%%%%%%%%%%%%%%%%%%%%%%%%%%%%%%%%%%%%%%%%%%%%%%
\section{Classical evolution}
                                       \label{sec:CE}

We consider {\it near flat} background, thus neglect $K$ term.
The equations and the large-scale solutions for the scalar- and tensor-type
structures can be written in a unified form as 
\bea
   & & {1 \over a^3 Q} (a^3 Q \dot \Phi)^\cdot 
       + c_A^2 {k^2 \over a^2} \Phi = 0,
   \label{Phi-eq}
\eea
where, for the fluid in Einstein gravity, the field in generalized gravity, 
and the tensor-type structures, respectively, we have
\cite{Hwang-Fluid-1999,HN-GGT-1996,Hwang-GW-1998}:
\bea
   & & \Phi = \varphi_v, \quad \; Q = { \mu + p \over c_A^2 H^2 }, 
       \qquad \qquad \qquad \quad \;\;\; c_A^2 = c_s^2
   \label{case-fluid} \\
   & & \Phi = \varphi_{\delta \phi}, \quad
       Q = { \omega \dot \phi^2 + {3 \dot F^2 \over 2 F}
       \over \left( H + {\dot F \over 2 F} \right)^2 } 
       \equiv {\dot \phi^2 \over H^2} Z_s,
       \quad c_A^2 = 1,
   \label{case-field} \\
   & & \Phi = C^\alpha_\beta, \quad \; Q = F \equiv {1 \over 8 \pi G} Z_t,
       \qquad \qquad \quad c_A^2 = 1,
   \label{case-GW}
\eea
where $Z$'s become unity in the limit of Einstein gravity\footnote{
         In the gravity with stringy correction terms
         \bea
            & & \xi(\phi) [ c_1 R_{GB}^2 + c_2 G^{ab} \phi_{;a} \phi_{;b}
                + c_3 \Box \phi \phi^{;a} \phi_{;a}
                + c_4 (\phi^{;a} \phi_{;a})^2 ],
            \\
            & & g(\phi) R \tilde R,
         \eea
         in the Lagrangian, where
         $R_{GB}^2 \equiv R^{abcd} R_{abcd} - 4 R^{ab} R_{ab} + R^2$ and
         $R \tilde R \equiv \eta^{abcd} R_{ab}^{\;\;\;\;ef} R_{cdef}$,
         we still have eq. (\ref{Phi-eq}) with more complicated
         $Q$ and $c_A^2$ \cite{Stringy}.
         }.
Equations (\ref{case-fluid},\ref{case-field}) are valid for 
single component fluid and field, whereas eq. (\ref{case-GW})
is valid in the presence of arbitrary numbers of fluid and field
as long as the tensor-type anisotropic stress vanishes.
The case of eq. (\ref{case-fluid}) is valid for an ideal fluid
in Einstein gravity with 
$c_s^2 \equiv \dot p / \dot \mu$\footnote{
         In the situation with general $K$, with \cite{Phi-history}
         \bea
            & & \Phi \equiv \varphi_v - {K/a^2 \over 4 \pi G (\mu + p)}
                \varphi_\chi,
            \label{Phi-K}
         \eea
         eqs. (\ref{Phi-eq},\ref{case-fluid}) are valid for an
         ideal fluid \cite{Hwang-Fluid-1999}, whereas the same equations with
         $c_A^2 \equiv 1 - 3 ( 1- c_s^2 ) K/k^2$ 
         (where $c_s^2$ is for the field) are valid
         for a minimally coupled scalar field \cite{HN-Fluids-2001}.
         }.
The case of eq. (\ref{case-field}) is valid for 
the second-order gravity system such as
either $f = F(\phi) R$ in the presence of a field $\phi$ or
$f = f(R)$ without the field.
The case of eq. (\ref{case-GW}) is valid for the general system
in eq. (\ref{action}).
Using $z \equiv a \sqrt{Q}$ and $v \equiv z \Phi$ eq. (\ref{Phi-eq}) becomes
\cite{Lukash-1980,Phi-history}
\bea
   & & v^{\prime\prime} + \left( c_A^2 k^2 - z^{\prime\prime} / z \right) v
       = 0.
   \label{v-eq}
\eea
In the large-scale limit, with $z^{\prime\prime} / z \gg c_A^2 k^2$, we have
an exact solution
\bea
   & & \Phi = C ({\bf x}) - D ({\bf x}) \int_0^t {dt \over a^3 Q}.
   \label{Phi-LS-sol}
\eea
Ignoring the transient solution we have a temporally conserved behavior
\bea
   & & \Phi ({\bf x}, t) = C ({\bf x}).
   \label{conservation}
\eea
{}For the scalar-type perturbation we can show that the decaying
solution in eq. (\ref{Phi-LS-sol}) is $({k \over aH})^2$ higher order
compared with the one in the zero-shear gauge \cite{HN-GGT-1996}.
Therefore, the non-transient solutions of $\Phi$
in the large-scale limit is generally {\it conserved}.
These conservation properties are valid considering
generally time varying $p (\mu)$, $V(\phi)$, $\omega(\phi)$,
and $f(\phi,R)$ [$F(\phi)$ for $\varphi_{\delta \phi}$ and
$f(R)$ for $\varphi_{\delta F}$],
thus are valid independently of changes in underlying gravity theory.
The unified analyses of the gravity theories belonging to
eq. (\ref{action}) are crucially important to make this point: that is,
since the solutions and the conservation properties are valid considering
general $p$, $V$, $\omega$, and $f$, we can claim that $\Phi$
remains conserved independently of changing equation of state,
field potential, and gravity sector.

%%%%%%%%%%%%%%%%%%%%%%%%%%%%%%%%%%%%%%%%%%%%%%%%%%%%%%%%%%%%%%
\section{Quantum generation}
                                                 \label{sec:QG}

We have shown that the growing solution of $\Phi$
is conserved in the large scale limit {\it independently}
of the specifics of the gravity theories
including changes between different gravity theories.
Thus, the classical evolution in the large scale is characterized by
the conserved quantity $C({\bf x})$ which encodes the information
about the spatial structure of the nontransient solution.
In order to have information about large scale structure,
we need the information about $\Phi = C({\bf x})$
which must have been generated from quantum fluctuations 
in the early inflationary stage of the universe;
gravity alone cannot generate the seed fluctuations out of the
spatially homogeneous and isotropic background.

We consider the quantum generation process in unified forms.
{}From eq. (\ref{Phi-eq}) we can construct the perturbed action 
in a unified form 
\cite{Lukash-1980,Mukhanov-1988,MFB-1992,Hwang-CT-1997,Hwang-GW-1998}
\bea
   & & \delta^2 S = {1 \over 2} \int a^3 Q \left( \dot \Phi^2
       - c_A^2 {1 \over a^2} \Phi^{|\gamma} \Phi_{,\gamma}
       \right) dt d^3 x.
   \label{perturbed-action}
\eea
This action as well as eqs. (\ref{Phi-eq},\ref{v-eq}) was first 
derived by Lukash in 1980 in the context of an ideal fluid \cite{Lukash-1980},
and later was derived in the context of a field \cite{Mukhanov-1988};
eq. (\ref{v-eq}) first appeared in the work by Field and Shepley in 1968 
\cite{Phi-history}.

In order to handle the quantum mechanical generations of the scalar-type
structure and the gravitational wave, we regard the perturbed parts
of the metric and matter variables as Hilbert space operators, 
$\hat \Phi ({\bf x}, t)$.
Since we are considering a flat three-space background, we may
expand $\hat \Phi$ in mode function expansion
\bea
   & & \hat \Phi ({\bf x}, t) = \int {d^3 k \over (2 \pi)^{3/2}}
       \left[ \hat a_k \Phi_k (t) e^{i {\bf k} \cdot {\bf x}}
       + \hat a_k^\dagger \Phi_k^* (t) e^{- i {\bf k} \cdot {\bf x}}
       \right],
   \nonumber \\
\eea
where $\Phi_k (t)$ is a mode function.
The annihilation and creation operators $\hat a_k$ and $\hat a_k^\dagger$
follow the standard commutation relations.
In the quantization process of the gravitational wave we need to take
into account of the two polarization states properly
\cite{Ford-Parker-1977,Hwang-GW-1998}.
{}From our perturbed action in eq. (\ref{perturbed-action}) we have
$\pi_\Phi \equiv {\partial {\cal L} \over \partial \dot \Phi}
= a^3 Q \dot \Phi$.
{}From the equal-time commutation relation
$[ \hat \Phi ({\bf x},t), \hat \pi_\Phi ({\bf x}^\prime, t)]
= i \delta^3 ({\bf x} - {\bf x}^\prime)$ we can derive
\bea
   \Phi_k \dot \Phi_k^* - \Phi_k^* \dot \Phi_k = {i / (a^3 Q)}.
   \label{commutation-unified}
\eea

Under the {\it ansatzs}\footnote{
       {}For solutions in the case of more general ansatz, 
       see \cite{Martin-Schwarz-2001}.}
\bea
   & & z^{\prime\prime}/z = n/\eta^2, \quad
       c_A^2 = {\rm constant},
   \label{ansatz}
\eea
the mode function has an exact solution
\bea
   & & \Phi_k (\eta) = {\sqrt{\pi |\eta|} \over 2 a \sqrt{Q}}
       \Big[ c_1 (k) H_\nu^{(1)} (x) 
       + c_2 (k) H_\nu^{(2)} (x) \Big],
   \label{Phi_k-sol}
\eea
where $\nu \equiv \sqrt{n + 1/4}$ and $x \equiv c_A k |\eta|$.
{}From the quantization condition we have 
\bea
   & & |c_2|^2 - |c_1|^2 = 1,
\eea
where for the gravitational wave this condition
should be met for each polarization state \cite{Hwang-GW-1998}.
The power spectrum based on the vacuum expectation value of $\hat \Phi$ is
\bea
   & & {\cal P}_{\hat \Phi} (k, \eta) 
       \equiv {k^3 \over 2 \pi^2} \int
       \langle \hat \Phi ({\bf x} + {\bf r}, t)
       \hat \Phi ({\bf x}, t) \rangle_{\rm vac} e^{-i {\bf k} \cdot {\bf r}}
       d^3 r
   \nonumber \\
   & & \quad
       = {k^3 \over 2 \pi^2} |\Phi_k (\eta)|^2.
\eea
{\it Assuming} the simplest vacuum state with $c_2 = 1$ and $c_1 = 0$
which corresponds to the flat spacetime quantum field theory vacuum state
with positive frequencies, in the large-scale limit we have\footnote{
        {}For $\nu = 0$ we have an additional
        $2 \ln{(c_A k |\eta|)}$ factor.
        }

\bea
   & & {\cal P}_{\hat \Phi}^{1/2} \Big|_{LS}
       = {H \over 2 \pi} {1 \over a H |\eta|} 
       {\Gamma(\nu) \over \Gamma(3/2)}
       \left( {k |\eta| \over 2} \right)^{3/2 - \nu}
       {1 \over c_A^\nu \sqrt{Q}},
   \label{P-LS} 
\eea
where we should consider additional $\sqrt{2}$ factor for the gravitational 
wave which follows from proper considering of the two polarization states
\cite{Hwang-GW-1998}.
We have 
${\cal P}_{\hat \Phi} |_{LS} = {\rm constant}$\footnote{
         Using $n \equiv q (q + 1)$ we can show $z \propto |\eta|^{-q}$
         and $\nu = q + {1 \over 2}$, thus
         ${\cal P}_{\hat \Phi} |_{LS} = {\rm constant}$ for $\nu > 0$;
         for $\nu = 0$ we additionally have 
         $z \propto \sqrt{|\eta|} \ln{|\eta|}$, thus 
         ${\cal P}_{\hat \Phi} |_{LS} = {\rm constant}$ as well.
         },
thus consistent with the general large-scale behavior in 
eq. (\ref{conservation}).
The spectral indices are
\bea
   & & n_S - 1 \equiv {d \ln {\cal P}_{\varphi_v} \over d \ln{k}}
       = 3 - 2 \nu_s, 
   \nonumber \\
   & & n_T \equiv {d \ln {\cal P}_{C^\alpha_\beta} \over d \ln{k}}
       = 3 - 2 \nu_t.
   \label{n-def}
\eea

%%%%%%%%%%%%%%%%%%%%%%%%%%%%%%%%%%%%%%%%%%%%%%%%%%%%%%%%%%%%%%
\section{Slow-roll inflation}
                                       \label{sec:SR}

We consider situations without the fluid, thus $c_A^2 = 1$.
We introduce the slow-roll parameters \cite{HN-GGT-1996}
\bea
   & & \epsilon_1 \equiv {\dot H \over H^2}, \quad
       \epsilon_2 \equiv {\ddot \phi \over H \dot \phi}, \quad
       \epsilon_3 \equiv {1 \over 2} {\dot F \over H F}, 
   \nonumber \\
    & & \epsilon_4 \equiv {1 \over 2} {\dot E \over H E}, \quad
        E \equiv F \left( \omega + {3 \dot F^2 \over 2 \dot \phi^2 F} \right).
\eea
Compared with the Einstein gravity in \cite{Stewart-Lyth-1993}
we have two additional parameters
$\epsilon_3$ and $\epsilon_4$ for the scalar-type perturbation which 
reflect the effects of additional parameters $F$ and $\omega$
in our generalized gravity; 
for the tensor-type perturbation we have one additional parameter 
$\epsilon_3$ from $F$.
{}From eqs. (\ref{case-field},\ref{case-GW}) we have
\cite{HN-GGT-1996,Hwang-GW-1998}
\bea
   & & {z_s^{\prime\prime} \over z_s} = a^2 \Big[
       H^2 \left( 1 - \epsilon_1 + \epsilon_2 - \epsilon_3 + \epsilon_4 \right)
       \left( 2 + \epsilon_2 - \epsilon_3 + \epsilon_4 \right)
   \nonumber \\
   & & \quad
       + H \left( - \dot \epsilon_1 + \dot \epsilon_2 - \dot \epsilon_3
       + \dot \epsilon_4 \right)
       - 2 ( {3 \over 2} - \epsilon_1 + \epsilon_2 - \epsilon_3
       + \epsilon_4 ) 
   \nonumber \\
   & & \quad
       \times H {\dot \epsilon_3 \over 1 + \epsilon_3}
       - {\ddot \epsilon_3 \over 1 + \epsilon_3}
       + 2 {\dot \epsilon_3^2 \over ( 1 + \epsilon_3)^2} \Big],
   \label{z_s} \\
   & & {z_t^{\prime\prime} \over z_t} = a^2 \left[
       H^2 \left( 1 + \epsilon_3 \right)
       \left( 2 + \epsilon_1 + \epsilon_3 \right)
       + H \dot \epsilon_3 \right],
   \label{z_t}
\eea
and $\int^\eta ( 1 + \epsilon_1 ) d \eta = - {1 / (aH)}$.

{\it Assuming} $\dot \epsilon_i = 0$  we have $(1 + \epsilon_1)\eta = -1/(aH)$, 
thus ansatzs in eq. (\ref{ansatz}) are satisfied with
\bea
   & & n_s = { ( 1 - \epsilon_1 + \epsilon_2 - \epsilon_3 + \epsilon_4 )
       ( 2 + \epsilon_2 - \epsilon_3 + \epsilon_4 )
       \over ( 1 + \epsilon_1 )^2 },
   \nonumber \\
   & & n_t = { ( 1 + \epsilon_3 ) ( 2 + \epsilon_1 + \epsilon_3 )
       \over (1 + \epsilon_1)^2 }.
   \label{n}
\eea
In such a case the rest of the exact results in \S \ref{sec:QG} are available. 
Since the large-scale structures are generated during short time
interval (about $60$ $e$-folds) of the latest inflation, we anticipate
time variation of $\epsilon_i$ during that period is negligible;
still this is an {\it assumption} we are making in the following.
Under this situation the power-spectra of the two-types of structures
in the large-scale limit are given in eq. (\ref{P-LS}) with the 
spectral indices given as
\bea
   & & n_S - 1 = 3 - \sqrt{4 n_s + 1}, \quad
       n_T = 3 - \sqrt{4 n_t + 1}.
   \label{indices-constant-n}
\eea
Thus, by imposing the condition of Zel'dovich specra 
($n_S -1 \simeq 0 \simeq n_T$) which is consistent with the
CMBR observation,
we can derive constraints on $\epsilon_i$'s, thus on the parameters
of the gravity theory ($V$, $\omega$, and $F$).

Now, to the first-order in the slow-roll parameters, i.e., 
further {\it assuming} $|\epsilon_i| \ll 1$,
from eq. (\ref{P-LS}) we can derive
\bea
   & & {\cal P}_{\hat \varphi_{\delta \phi}}^{1/2} \Big|_{LS}
       = {H \over |\dot \phi|} {\cal P}_{\delta \hat \phi_\varphi}^{1/2} 
       \Big|_{LS}
       = {H^2 \over 2 \pi |\dot \phi|} {1 \over \sqrt{Z_s}}
       \Big\{ 1 + \epsilon_1 
   \nonumber \\
   & & \quad
       + \big[ \gamma_1 + \ln{(k |\eta|)} \big] 
       ( 2 \epsilon_1 - \epsilon_2 + \epsilon_3 - \epsilon_4 ) \Big\},
   \label{P-SR-scalar} \\
   & & {\cal P}_{\hat C^\alpha_\beta}^{1/2} \Big|_{LS}
       = \sqrt{16 \pi G} {H \over 2 \pi} {1 \over \sqrt{Z_t}}
   \nonumber \\
   & & \quad
       \times \Big\{ 1 + \epsilon_1 
       + \big[ \gamma_1 + \ln{(k |\eta|)} \big] 
       ( \epsilon_1 - \epsilon_3 ) \Big\},
   \label{P-SR-tensor}
\eea
where $\gamma_1 \equiv \gamma_E + \ln{2} - 2 = - 0.7296 \dots$, 
with $\gamma_E$ the Euler constant.
We have [we have corrected an error in $Z_s$ in the published version]
\bea
   & & Z_s = { E/F \over (1 + \epsilon_3)^2}, \quad
       Z_t = 8 \pi G F.
\eea
Thus, besides $\epsilon_1$, 
the scalar-type perturbation is affected by 
$\epsilon_2$, $\epsilon_3$ and $\epsilon_4$
(thus, $\phi$, $F$ and $\omega$), whereas the tensor-type perturbation
is affected by $\epsilon_3$ (thus, $F$) only; see also eq. (\ref{indices}).

The observationally relevant scales exit Hubble horizon within about
$60$ $e$-folds before the end of the latest inflation.
{}Far outside the horizon the quantum fluctuations classicalize
and we can identify ${\cal P}_{\Phi} = {\cal P}_{\hat \Phi}$ where
${\cal P}_{\Phi}$ is the power-spectrum based on spatial averaging
\bea
   & & {\cal P}_{\Phi} (k, \eta) 
       \equiv {k^3 \over 2 \pi^2} \int
       \langle \Phi ({\bf x} + {\bf r}, t)
       \Phi ({\bf x}, t) \rangle_{\bf x} e^{-i {\bf k} \cdot {\bf r}} d^3 r
   \nonumber \\
   & & \quad
       = {k^3 \over 2 \pi^2} |\Phi (k,\eta)|^2,
\eea
with $\Phi (k,\eta)$ a Fourier transform of $\Phi ({\bf x}, \eta)$. 
Since $\Phi$ is conserved in the large-scale limit,
the power-spectra in eqs. (\ref{P-SR-scalar},\ref{P-SR-tensor})
can be {\it identified} as the classical power-spectra at later epoch.
We have in mind a scenario in which the inflation based on a field or
a generalized gravity is followed by ordinary radiation and matter 
dominated eras based on Einstein gravity.
We have shown in eq. (\ref{conservation}) that as long as the scale 
remains in the super-horizon scale $\Phi$ is conserved independently 
of the changing gravity theory from one type to the other.
Therefore, eqs. (\ref{P-SR-scalar},\ref{P-SR-tensor}) are now valid for
the classical power-spectra.
The spectral indices of the scalar and tensor-type perturbations
in eq. (\ref{n-def}) become
\bea
   & & n_S - 1 = 2 ( 2 \epsilon_1 - \epsilon_2 + \epsilon_3 - \epsilon_4 ),
       \quad
       n_T = 2 ( \epsilon_1 - \epsilon_3 ).
   \label{indices}
\eea

{}For the scale independent Zel'dovich ($n_S -1 \simeq 0 \simeq n_T$) spectra
the quadrupole anisotropy becomes 
\bea
   & & \langle a_2^2 \rangle
       = \langle a_2^2 \rangle_S + \langle a_2^2 \rangle_T
       = {\pi \over 75} {\cal P}_{\varphi_{\delta \phi}}
       + 7.74 {1 \over 5} {3 \over 32} {\cal P}_{C_{\alpha\beta}},
   \label{a_2}
\eea
which is valid for $K = 0 = \Lambda$; 
for a general situation with nonvanishing $\Lambda$, see \cite{Knox-1995}.
The four-year {\it COBE}-DMR data give 
$\langle a_2^2 \rangle \simeq 1.1 \times 10^{-10}$, \cite{COBE}.
{}From eqs. (\ref{P-SR-scalar},\ref{P-SR-tensor}) 
the ratio between two types of perturbations 
$r_2 \equiv {\langle a_2^2 \rangle_T / \langle a_2^2 \rangle_S}$ becomes
[we have corrected an error in eq. (\ref{ratio})
and the rest of this paragraph in the published 
version\footnote{
       We thank David Wands for pointing out a possible error
       in eq. (\ref{ratio}) of the previous version based on the
       conformal transformation argument.

       Using the conformal transformation properties 
       in eqs. (10-14) of \cite{Hwang-CT-1997} we can show that
       the slow-roll parameters transform 
       to the linear order as
       \bea
          & & \hat \epsilon_1 = \epsilon_1 - \epsilon_3, \quad
              \hat \epsilon_2 = \epsilon_2 - 3 \epsilon_3 + \epsilon_4,
          \nonumber 
       \eea
       where hats denote quantities in the conformally transformed
       Einstein frame.
       Thus, we have 
       \bea
          & & \hat n_S - 1 = 2 ( 2 \hat \epsilon_1 - \hat \epsilon_2 )
              = 2 ( 2 \epsilon_1 - \epsilon_2 + \epsilon_3 - \epsilon_4 )
              = n_S - 1,
          \nonumber \\
          & & \hat n_T = 2 \hat \epsilon_1 = 2 (\epsilon_1 - \epsilon_3) = n_T,
          \nonumber \\
          & & \hat r_2 = | - 13.8 \hat \epsilon_1 |
              = | - 13.8 (\epsilon_1 - \epsilon_3) | = r_2.
          \nonumber 
       \eea
       These results are consistent with the fact that
       $\varphi_{\delta \phi}$ and $C_{\alpha\beta}$
       are invariant under the conformal transformation
       as shown in eq. (25) of \cite{Hwang-CT-1997}.

       In fact, using the conformal transformation properties
       in eqs. (10-14,25) of \cite{Hwang-CT-1997}
       we can check that eq. (\ref{Phi-eq}) and
       eq. (\ref{perturbed-action}), thus all the consequent
       results in the classical evolution and the quantum
       generation, can be simply drived from the known results
       in the Einstein frame.
       This does not mean that the results in the two (the original
       and the Einstein frames) are equivalent.
       It only means that the results in the two frames are
       related through the conformal transformation.
       }]
\bea
   r_2 
   &=& 13.8 \times 4 \pi G {\dot \phi^2 \over H^2} \left| {Z_s \over Z_t}
       \right|
   \nonumber \\
   &=& 13.8 {1 \over ( 1 + \epsilon_3 )^2}
       \left| {\omega \dot \phi^2 \over 2 H^2 F} + 3 \epsilon_3^2 \right|
   \nonumber \\
   &=& \left| - 13.8 {1 \over ( 1 + \epsilon_3 )^2}
       \left[ (\epsilon_1 - \epsilon_3) ( 1 + \epsilon_3)
       - {\dot \epsilon_3 \over H} \right] \right|
   \nonumber \\
   &=& \left| - 13.8 (\epsilon_1 - \epsilon_3) \right|,
   \label{ratio}
\eea
where in the last step we used the linear slow-roll condition.
In the limit of Einstein gravity we have $r_2 = - 13.8 \epsilon_1
= - 6.92 n_T$ which is independent of $V$ and is
the well known consistency relation; $\epsilon_1$ is always negative
in the ordinary slow-roll inflation, thus $n_T$ is negative.
Even in our class of generalized gravity theories,
from eqs. (\ref{indices},\ref{ratio}) we have
$r_2 = \left| - 13.8 (\epsilon_1 - \epsilon_3) \right| 
= \left| - 6.92 n_T \right|$,
thus the consistency relation (in the amplitude!)  remains valid.
However, in our generalized gravity case, 
$n_T$ in eq. (\ref{indices}) could have either sign depending on situations.

%%%%%%%%%%%%%%%%%%%%%%%%%%%%%%%%%%%%%%%%%%%%%%%%%%%%%%%%%%%%%%
%\section{Applications}
%                                       \label{sec:Applications}

Inflation based on Einstein gravity with a minimally coupled scalar field
is a simple case with $F = 1/(8 \pi G)$ and $\omega = 1$.
In this case we have $\epsilon_3 = 0 = \epsilon_4$ and $Z = 1$.
The power spectra of slow-roll inflation \cite{Stewart-Lyth-1993}
belong to eqs. (\ref{P-SR-scalar},\ref{P-SR-tensor},\ref{indices}).
Accuracy of the slow-roll approximation compared with the exact
integration of the fundamental equation in eq. (\ref{Phi-eq})
has been discussed in \cite{Slow-roll}.
{}For a recent attempt to consider higher-order effects of the
slow-roll parameters in a perturbative approach, 
thus going beyond the ansatz made in eq. (\ref{ansatz}), 
see \cite{Stewart-Gong-2001}.

The presence of $F$ and $\omega$, thus $\epsilon_3$, $\epsilon_4$ and $Z$'s in 
eqs. (\ref{P-SR-scalar},\ref{P-SR-tensor},\ref{indices},\ref{ratio}) 
indicates the deviation from the Einstein gravity.
Inflationary spectra in various inflationary models based 
on generalized gravity theories made in \cite{HN-GGT-infl,HN-fR-2001}
can be recovered by simply reducing our general results in this paper.

%%%%%%%%%%%%%%%%%%%%%%%%%%%%%%%%%%%%%%%%%%%%%%%%%%%%%%%%%%%%%%
\section{Discussions}
                                       \label{sec:Discussions}

As we have shown in this paper, even in a class of generalized gravity
theories included in eq. (\ref{action}) we can present the results
quite similarly as in Einstein gravity case and in unified forms.
The effects of generalized gravity appear in two additional parameters
$F$ and $\omega$ which are reflected in the two additional slow-roll
parameters $\epsilon_3$ and $\epsilon_4$.
One important underlying reason for such simple results in 
apparently complicated and diverse gravity theories belonging to
eq. (\ref{action}) can be traced to the conformal transformation
property of the gravity theory we are considering \cite{CT}.

In addition to the coherent and unified presentation of the
classical evolution and quantum generation processes,
the slow-roll power spectra in \S \ref{sec:SR} can be
regarded as new contributions of the present work.
These results, in fact, include results from most of the inflationary
scenarios based on generalized gravity as well as Einstein gravity theory.
In generic forms, eqs. (\ref{P-SR-scalar},\ref{P-SR-tensor},\ref{indices}) 
show the amplitudes and spectral indices of the generated structures, 
and eq. (\ref{ratio}) shows the ratio of gravitational wave contribution 
relative to the scalar-type structure.
Various previous studies on the subject can be regarded as specific
limits of these generic results.

%%%%%%%%%%%%%%%%%%%%%%%%%%%%%%%%%%%%%%%%%%%%%%%%%%%%%%%%%%%%%%
\section*{Acknowledgments}

We thank Cyril Cartier for useful information and
Ewan Stewart for useful comments.
We also wish to thank Dominik Schwarz for useful discussions and 
insightful comments.
JH was supported by the Korea Research Foundation Grants
(KRF-2000-013-DA004 and 2000-015-DP0080).
HN was supported by grant No. 2000-0-113-001-3 from the
Basic Research Program of the Korea Science and Engineering Foundation.

%%%%%%%%%%%%%%%%%%%%%%%%%%%%%%%%%%%%%%%%%%%%%%%%%%%%%%%%%%%%%%

%%%%%%%%%%%%%%%%%%%%%%%%%%%%%%%%%%%%%%%%%%%%%%%%%%%%%%%%%%%%%%
\end{document}